# On-chip amplification-free $f_{CEO}$ detection and broadband SCG in parabolically width-modulated TFLN waveguides

TIANYOU TANG,[1] SIMIN YU,[2] RUIXIN ZHOU,[2] JIANING ZHANG,[3] JUANJUAN LU,[2, *] GUANYU CHEN,[3, *] TAO ZHU,[3] AND LINGFANG WANG[1, *]

[1]*School of Optoelectronic Science and Engineering, University of Electronic Science and Technology of China, Chengdu 611731, Sichuan, P. R. China*
[2]*School of Information Science and Technology, ShanghaiTech University, Shanghai 201210, P. R. China*
[3]*Key Laboratory of Optoelectronic Technology & Systems (Ministry of Education), Chongqing University, Chongqing 400044, P. R. China*
*Corresponding authors：Lf.wang@uestc.edu.cn, lujj2@shanghaitech.edu.cn, gychen@cqu.edu.cn

**Abstract:** We demonstrate amplification-free carrier-envelope offset ($f_{CEO}$) detection in parabolically width-modulated z-cut thin-film lithium niobate (TFLN) waveguide that simultaneously engineers dispersion and enhances nonlinear interactions. With 116 fs pulses at 1560 nm, the device generates more than two octaves broadband SCG spanning 600 to beyond 2400 nm, enabling spectral overlap between dispersive waves and second-harmonic generation near 780 nm for robust $f$–$2f$ self-referencing. Most notably, we achieve the first fully amplification-free, on-chip $f_{CEO}$ detection without optical or electronic amplifiers, obtaining 34 dB SNR at 100 kHz RBW and 57 dB at 100 Hz RBW, the highest reported to date under such conditions. These results establish a new benchmark for scalable, power-efficient, self-referenced frequency combs and quantum-enabled photonic systems.

## 1. Introduction

Optical frequency combs (OFCs) have revolutionized timekeeping [1-6], spectroscopy [7-11], and precision metrology [12-14] by providing a phase-coherent bridge between microwave and optical frequencies [15-17]. At the heart of comb stabilization lies the detection of the carrier-envelope offset frequency ($f_{CEO}$) via $f$–$2f$ self-referencing, which requires an octave-spanning supercontinuum and efficient second-harmonic generation (SHG). Traditionally, such systems rely on bulky optics and free-space interferometers, posing challenges for alignment, power consumption, and integration [18-22]. Recent efforts in integrated photonics have sought to miniaturize this functionality, enabling chip-scale frequency combs for portable atomic clocks, precision LIDAR, and quantum networks [23-35].

Among emerging integrated platforms, thin-film lithium niobate (TFLN) has attracted significant attention due to its simultaneous support for high $\chi^{(2)}$ and $\chi^{(3)}$ nonlinearities, low optical losses, and mature fabrication processes [36]. Recent demonstrations have leveraged TFLN waveguides to achieve octave-spanning SCG [37-40] and on-chip $f$–$2f$ interferometry [38-40]. However, most existing implementations employ uniform or linear-tapered geometries [37-40], limiting dispersion engineering flexibility and constraining phase-matching bandwidth. To overcome these limitations, we propose and implement a *z-cut* TFLN waveguide with a parabolically modulated width profile shown in Fig. 1(a). This structure enables adiabatic modulation of group velocity dispersion (GVD) and nonlinear interaction strength along the propagation axis, supporting efficient soliton dynamics, dispersive wave (DW) generation, and SHG-SCG spectral overlap. We experimentally demonstrate >2-octave SCG and $f_{CEO}$ detection of ~34 dB SNR at 100 kHz RBW and 57dB SNR at 100 Hz RBW, without external amplification and external filtering. This work establishes a robust, power-efficient solution for fully integrated comb stabilization.

Critically, we extend prior work by systematically investigating the dual roles of pump power and input polarization in fine-tuning SCG dynamics. Pump power governs the balance between self-phase modulation and soliton fission, directly impacting DW emissions, while input polarization dictates $\chi^{(2)}$-mediated SHG efficiency and $\chi^{(3)}$-driven spectral broadening. This work not only resolves the long-standing trade-off between SCG bandwidth and nonlinear efficiency in integrated platforms but also establishes a robust framework for chip-scale optical atomic clocks. Our findings highlight the untapped potential of dispersion-engineered TFLN waveguides in advancing portable precision metrology and distributed quantum sensing networks.

## 2. Nonlinear Dynamics in Parabolically Modulated Waveguides

In nonlinear integrated photonics, the GVD plays a central role in shaping the propagation dynamics of femtosecond pulses. For conventional waveguides with uniform cross-sections, the GVD parameter is constant along the propagation axis and primarily determined by the waveguide geometry and refractive index contrast. This fixed dispersion profile limits the available phase-matching bandwidth and spectral overlap between interacting waves of the fundamental and second-harmonic components.

In contrast, our parabolically tapered *z-cut* TFLN waveguide features a continuously varying width from 2.0 to 0.6 μm over a 6 mm propagation length (Fig. 1(a)). This structural profile dynamically alters the effective refractive index $n_{eff}(\omega, z)$ which varies with the position $z$ along the waveguides' axial direction, and thus the propagation coefficient $\beta(\omega, z)$. Consequently, the second-order dispersion parameter becomes position-dependent, as expressed by equation (1), where $\lambda$ and $c$ denote wavelength and the speed of light, respectively.

$$D = -\frac{\lambda}{c}\frac{d^2 n_{eff}(\omega,z)}{d\lambda^2} \qquad (1)$$

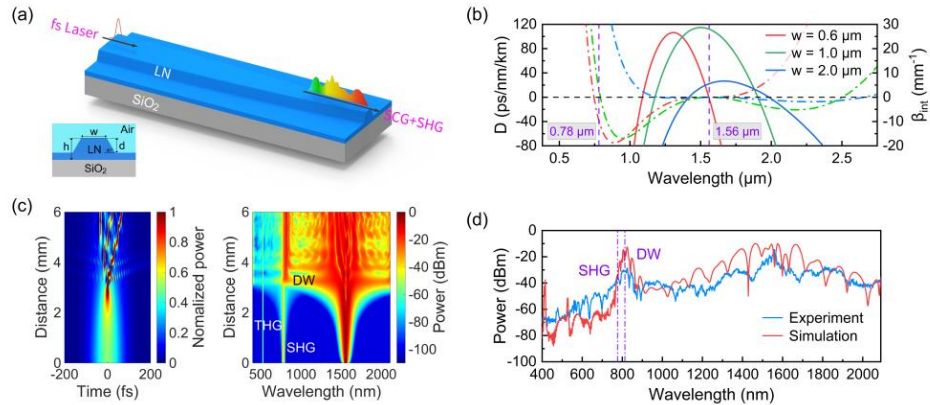

**Fig.1.** (a) Illustration of the width-modulated TFLN waveguide and its cross-sectional structure. (b) GVD calculations for waveguide widths varying from 2 μm at input facet to output end width of 0.6 μm (solid lines) and calculations of $\beta_{int}$ of different waveguide widths of 2 μm, 1 μm and 0.6 μm (dashed lines), which indicate the DW phase matched wavelength at ~780 nm. (c) Simulated temporal (left) and spectral (right) evolution in a 6 mm long with modulated TFLN waveguide based on modified GNLSE Equation (2). (d) SCG simulation compared with experimental result at the same pumping condition.

Figure 1(b) shows the simulated GVD as a function of wavelength for various waveguide widths, corresponding to different positions along the parabolically tapered structure. As the waveguide width decreases continuously from 2.0 μm at the input to 0.6 μm at the output, the GVD for the fundamental TE mode evolves from anomalous dispersion to normal dispersion. This spatially varying dispersion enables a dynamic sequence of nonlinear processes: soliton fission occurs near the input region under anomalous dispersion, while DW emission emerges near the output in the normal dispersion regime, particularly around 780 nm which was predicted in the calculation in Fig. 1(b) based on equation (2), where $\beta_\omega$ denotes the propagation constant at frequency $\omega$. The pump frequency is given by $\omega_0$, and $v_g$ corresponds to the group velocity at $\omega_0$.

$$\beta_{int}(\omega) = \beta_\omega - \beta_{\omega_0} - \frac{1}{v_g}(\omega - \omega_0) = 0 \tag{2}$$

To model the femtosecond pulse propagation through this parabolic width modulated waveguide, we modified a generalized nonlinear Schrödinger equation (GNLSE) that includes both second- and third-order nonlinearities, higher-order dispersion, self-steepening, and linear loss, which are all modulated by width along the propagation axis, represented in equation (3). Here, $A(z, t)$ is the complex envelope of the optical field; $\beta_k(z)$ are the dispersion coefficients varying with propagation position $z$; $\alpha(z)$ is the linear loss coefficient; $\gamma(z)$ is the effective third-order nonlinear coefficient given by $\gamma(z) = \omega_0 n_2/(c\, A_{eff}(z))$, where $A_{eff}(z)$ is the effective mode area. The self-steepening term is included through the temporal derivative $\partial A/\partial z$ scaled by the inverse pump frequency $1/\omega_0$. The $\chi^{(2)}$ contribution is modeled by the effective coefficient $\kappa(z) = (\omega_0 d_{eff}/cn_0)\zeta(z)$, where $\zeta(z)$ is the normalized mode overlap integral between the pump and SHG modes.

$$\frac{\partial A(z,t)}{\partial z} + \frac{\alpha(z)}{2} A(z,t) - \sum_{k\geq 2} \frac{i^{k+1}}{k!} \beta_k(z) \frac{\partial^k A(z,t)}{\partial t^k} = i\gamma(z)\left(1 + \frac{i}{\omega_0}\frac{\partial}{\partial t}\right)|A|^2 A + i\kappa(z)A^2 \tag{3}$$

To further elucidate the nonlinear dynamics underlying the observed broadband supercontinuum generation, we numerically modeled pulse propagation in the parabolically tapered TFLN waveguide using the modified GNLSE, incorporating both second- and third-order nonlinearities, high-order dispersion, and self-steepening effects. The simulated results are presented in Fig. 1(c), showing both the temporal evolution (left panel) and spectral evolution (right panel) of the optical pulse along the 6-mm-long taper.

In the temporal domain, the pulse initially undergoes self-phase modulation and temporal compression, followed by soliton fission and the generation of sub-pulses at varying group velocities. These features indicate soliton dynamics mediated by the evolving GVD from anomalous to near normal regimes along the taper. While in the spectral domain, the initial broadening is dominated by SPM around the pump (~1560 nm), followed by strong spectral extension into the visible and mid-infrared regions. Notably, distinct spectral features emerge ~ 780 nm and ~500 nm, corresponding to the generation of phase-matched DW emissions, SHG, and third-harmonic generation (THG), respectively. The visible DW component near 780 nm spectrally overlaps with the SHG signal, supporting the experimental observation of coherent *f*–2*f* beatnote generation.

The simulated temporal and spectral dynamics in Fig. 1(c) not only reveal the complex nonlinear evolution within the tapered TFLN waveguide but also reproduce key experimental signatures observed in broadband SCG. To validate the fidelity of these simulations, we directly compare the output spectrum from simulation with experimental results at 293 pJ on-chip energy, as shown in Fig. 1(d).

The comparison confirms excellent agreement between theory and experiment, particularly in the visible region near 780 nm, where both the SHG and phase-matched DW components emerge. The spectral overlap between these two features—highlighted by the dashed window—represents a critical condition for *f*–2*f* interferometry. Additionally, both spectra exhibit similar bandwidths and spectral slopes across the near-infrared region, indicating that the modeled dispersion profile and nonlinear coefficients accurately capture the essential physics of the parabolically tapered waveguide. This agreement reinforces the conclusion that the dynamic GVD modulation—transitioning from anomalous to normal dispersion along the waveguide—enables simultaneous soliton-driven DW emission and efficient SHG. The resulting spectral overlap is not only predictable but tunable, providing a robust and scalable pathway for chip-based self-referenced frequency comb systems.

## 3. Experimental Methods and Characterization

The waveguides were fabricated on a 600 nm thick z-*cut* TFLN bonded on a 2 μm $SiO_2$ buffer layer and a silicon substrate. The waveguide geometry was defined using electron-beam lithography (EBL) followed by reactive ion etching (RIE) to transfer the pattern. Two types of ridge waveguides were fabricated for comparison: a straight waveguide with fixed width of 1.0 μm, and a parabolically tapered waveguide with width varying smoothly from 2.0 to 0.6 μm over a 6 mm propagation length, with etching height of 400 nm, and etching angle of ~60°.

A schematic of the experimental setup is shown in Fig. 2(a). A mode-locked fiber laser generated 116 *f*s pulses which was polarized in TM mode at 1560 nm with an 80 MHz repetition rate, delivering 195 pJ pulse energy (~100 mW average power) to the waveguide input through a polarization-maintaining lensed fiber. A half-wave plate was placed after the laser beam to tune the input polarization. Total coupling loss was estimated to be ~16 dB. The output light from the TFLN waveguide was collimated using an aspheric lens (NA = 0.6) and splitted into two paths: one directed to optical spectrum analyzers (ANDO AQ6315A, 350-1750 nm, and Yokogawa AQ6375B, 1200-2400 nm) for broadband SCG characterizations, and the other focused onto a photodiode to detect the $f_{CEO}$ beatnote arising from interference between the DW and overlapped.

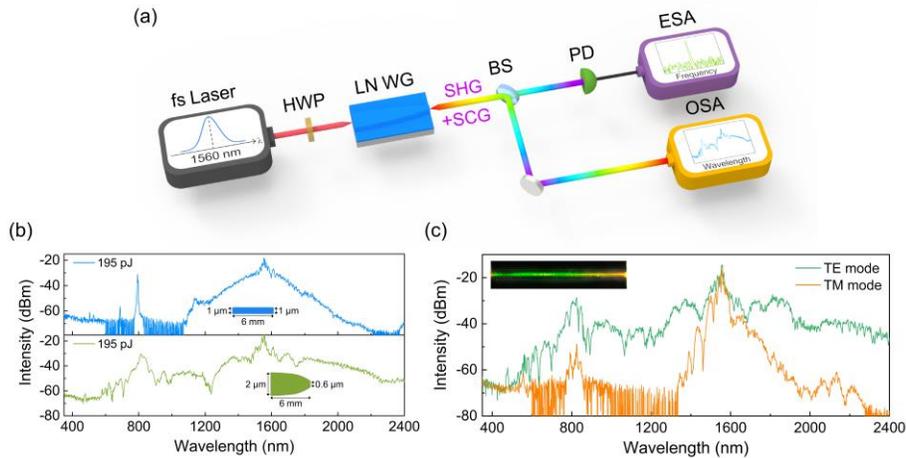

**Fig.2.** (a) Schematic experimental setup for SCG and $f_{CEO}$ characterizations. (b) Comparison of SCG generated in uniform width (top) and parabolic width-modulated (bottom) TFLN waveguides under identical pump conditions (1560 nm, 116 fs, 195 pJ). The insets depict the corresponding waveguide width profiles over a propagation length of 6 mm. (c) Polarization-dependent supercontinuum spectra generated in a parabolic width-modulated TFLN waveguide under the same pulse energy ~293 pJ. The TE mode (green) exhibits broad spectral extension with strong DW emission near 800 nm, while the TM mode (orange) results in significantly reduced bandwidth due to lower effective nonlinearities and phase-mismatch for SHG and DW processes. The inset is a photograph of the scattered light taken from above the 6 mm long chip

The SCG behavior was first compared between the uniform-width and parabolically tapered waveguides under identical pump conditions:195 pJ on-chip pulse energy and TE polarization. As shown in Fig. 2(b), the width modulated waveguide exhibited a significantly broader SCG spectrum spanning from 600 nm to beyond 2400 nm, compared to the uniform waveguide which generated a limited 1-octave bandwidth (1100–2200 nm). Beyond extending the bandwidth from 600 nm to beyond 2400 nm, the parabolic taper uniquely produces broadband and power-scalable spectral overlap between the SHG and dispersive wave in z-cut TFLN, which was not realized in prior work. This overlap not only enhances the robustness of *f*–2*f* self-referencing, but also opens new routes for advanced schemes such as *f*–3*f* referencing, multi-harmonic mixing, and quantum photonics requiring coherent visible-to-infrared coverage.

To further investigate the polarization dependence, we measured SCG spectra in the same waveguide under TM-polarized pumping. As displayed in Fig. 2(c), the bandwidth was significantly reduced (~1300–2200 nm) and no visible DW emission was observed. This confirms the essential role of TE-mode confinement and phase-matching in enabling efficient nonlinear interactions in *z-cut* TFLN.

Focus on the parabolically width- modulated TFLN waveguide, we proceeded to examine how the SCG in this waveguide evolves with varying pump pulse energy. As shown in Fig. 3, SCG spectra were measured across a range of on-chip power of ~1.6-23.4 mW (corresponding to ~20-293 pJ on chip pulse energy). A clear nonlinear threshold appeared near ~11 mW(~138 pJ), above which rapid spectral broadening and DW formation occurred. Notably, both the SHG component near 780 nm and the DW shoulder around 780 nm intensified with increasing energy, eventually blending into a broad visible output. This power-dependent evolution suggests that soliton dynamics and phase-matched DW emission are strongly coupled to the emergence of SHG, creating ideal conditions for coherent *f*–2*f* mixing.

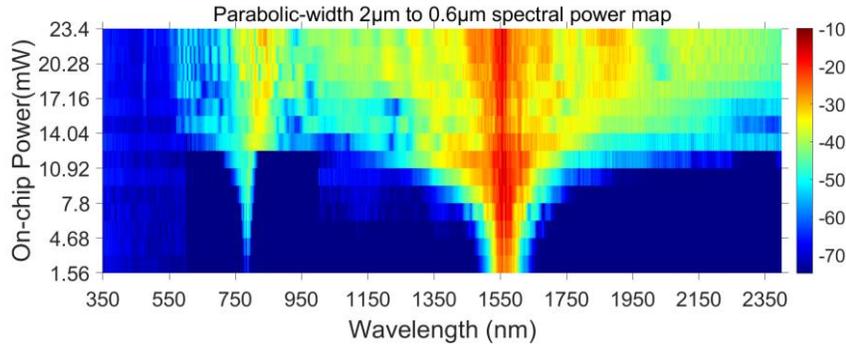

**Fig.3.** Experimental spectral evolution as a function of on-chip power for the parabolically tapered TFLN waveguide. The color map reveals the power-dependent supercontinuum generation dynamics across the 350–2400 nm range, highlighting the emergence of DW radiation around 780 nm and broadband extension into the mid-infrared. The spectral intensity is plotted in logarithmic scale with units of dBm.

We then examined how the SCG spectrum evolves under different input polarization angles in the tapered structure. The resulting spectral map, displayed in Fig. 4, TE-dominant input (0°–30°) supporting broadband SCG and strong visible DWs. This favorable dispersion-nonlinearity interplay facilitates soliton fission formation and subsequent DW radiation near 780 nm, as well as the generation of a coherent SHG component in the same spectral region. TM-dominant input (>60°) resulting in compressed spectra and suppressed SHG/DW features. This behavior arises from the TM mode inhibiting soliton dynamics and suppressing dispersive wave generation. The optimal polarization range (0°–30°) corresponds to the point where DW and SHG strongly overlap—critical for coherent *f*–2*f* detection.

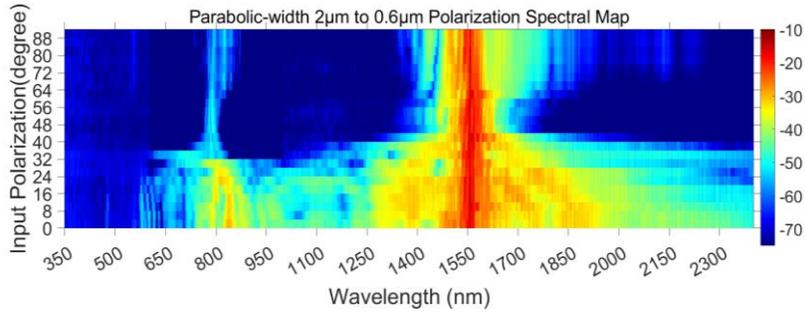

**Fig.4.** Experimental spectral evolution as a function of input polarization angle for the parabolically tapered TFLN waveguide. The color map reveals the strong polarization dependence of supercontinuum generation, highlighting the dominant role of TE-like input polarization in enabling efficient spectral broadening, dispersive wave emission (~780 nm), and second-harmonic generation. The spectral intensity is plotted in logarithmic scale with units of dBm.

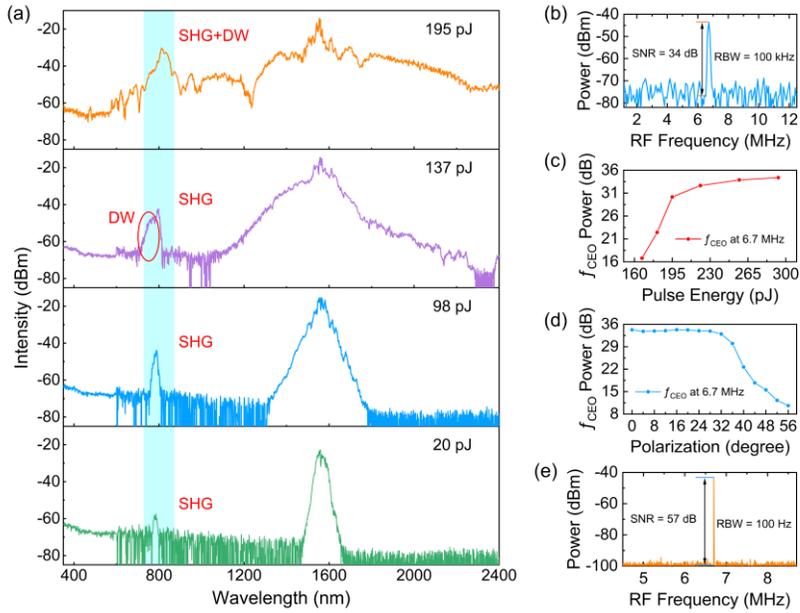

**Fig.5.** (a) Output spectra from the parabolic-width TFLN waveguide at increasing pulse energies (20–195 pJ), showing progressive development of SHG and dispersive wave (DW) features near 780 nm and broadband SCG beyond 2400 nm. At 137 pJ, a DW signal of comparable amplitude appears adjacent to the SHG peak, which merges into a broad visible band at 195 pJ. (b) Detected $f_{CEO}$ beatnote at 100 kHz RBW with 34 dB SNR, without any amplification. (c) $f_{CEO}$ power scaling with increasing pulse energy, demonstrating the threshold for coherent $f$–$2f$ self-referencing. (d) $f_{CEO}$ power as a function of input polarization, indicating peak performance under TE-like polarization. (e) $f_{CEO}$ beatnote measured at 100 Hz RBW, exhibiting a 57 dB SNR—the highest reported SNR in amplification-free detection to date.

To explore this mechanism further, we extracted representative spectra at selected pulse energies from 20 to 195 pJ under optimized polarization, as shown in Fig. 5(a). At low energy (20 pJ), only a weak SHG peak is observed. As energy increases to 137 pJ, a phased-matched DW emerges to the SHG band, and by 195 pJ, the SHG and DW merge totally into a single broad visible band. The co-evolution facilitates coherent $f$-$2f$ mixing.

This is confirmed by $f_{CEO}$ detection shown in Fig. 5(b), with a 34 dB SNR at 100 kHz RBW, without any optical amplification or RF boosting. Importantly, as shown in Fig.5(e), when using a 100 Hz RBW, the SNR further increased to ~57 dB. To the best of our knowledge, this is the first demonstration of $f_{CEO}$ detection entirely without optical amplification, APDs, or RF amplifiers, reaching 57 dB SNR at 100 Hz RBW. Previous reports required electronic amplification, whereas our fully amplification-free approach establishes a new benchmark for integration-ready and power-efficient frequency comb stabilization. This validates the strong phase coherence between SHG and DW components. Fig.5(c) and 5(d) further demonstrate the scalability and polarization dependence of the $f_{CEO}$ signal. The beatnote power increases with pump energy and peaks under TE-dominant input (0°–30°), in line with enhanced $\chi^{(2)}$ and $\chi^{(3)}$ efficiency. The signal remains stable even at higher powers, underscoring the robustness of the $f$-$2f$ scheme in this architecture. The exceptional 57 dB SNR achieved at 100 Hz RBW without any amplification establishes a new benchmark in integrated $f_{CEO}$ detection and highlights the efficacy of DW-SHG spectral overlap engineered by parabolic tapering.

## 4. Discussion

To further highlight the novelty and advantages of our work, **Table 1** presents a comparative summary of recent demonstrations of $f_{CEO}$ detection across various material platforms and waveguide geometries. Unlike previous reports that relied on either optical amplification or electronic boosting devices—such as avalanche photodiodes (APDs) with internal gain or RF power amplifiers—to extract a detectable $f_{CEO}$ signal, our approach achieves fully amplification-free detection of 34 dB SNR at 100 kHz RBW and an unprecedented 57 dB SNR at 100 Hz RBW—representing the highest SNR reported to date under entirely amplification-free conditions. This distinction is critical for realizing compact and robust on-chip systems that can directly interface with low-noise electronics.

**Table 1. Comparison between this work with literatures.**

(FW: Fixed-width, PWM: Parabolically width-modulated, GF: Gas-filled, HCF: Hollow capillary fiber, ML: Metalens, SCF: Suspended core fiber, EM: Evolution map.)

| Configuration | Material | Power EM | Polarization EM | $f_{CEO}$ Power | With Amplification | Ref. |
|---|---|---|---|---|---|---|
| FW-WG | LN (x-*cut*) | No | No | 30 dB (@1 MHZ) | Yes (APD ~34dB at 800nm) | [38] |
| FW-WG | LN (x-*cut*) | No | No | 50 dB (@1 Hz) | Yes (APD ~34dB at 800nm) | [39] |
| FW-WG | LN (x-*cut*) | No | No | 56 dB (@30 kHz) | Yes (RF power amplifier) | [40] |
| GF-HCF | Silica-glass | Yes | No | None | No | [41] |
| ML-SCF | $SiO_2$ | Yes | No | None | No | [42] |
| PWM WG | LN (z-*cut*) | Yes | Yes | 34/42/57 dB (@100 kHz /30 kHz/100 Hz) | No | **This work** |

Moreover, while most prior demonstrations are based on x-cut TFLN waveguides and support only limited polarization states, our z-cut platform enables simultaneous optimization of both $\chi^{(2)}$ and $\chi^{(3)}$ nonlinearities under TE excitation. This dual enhancement not only facilitates broadband SCG at lower pulse energies but also ensures spectral overlap between the SHG and DW components—critical for coherent $f$–$2f$ self-referencing.

Together, these features offer a compact, power-efficient, and highly scalable solution for chip-scale optical frequency comb stabilization, positioning our design as a significant step forward in integrated nonlinear photonics.

## 5. Conclusion

In conclusion, we have demonstrated a dispersion-engineered strategy for on-chip SCG and self-referenced frequency combs using a parabolically tapered z-*cut* TFLN waveguide. Compared to uniform-width designs, the parabolic taper enables dynamic modulation of group velocity dispersion, which promotes efficient soliton fission and phase-matched DW emission near the SHG band. This spatial dispersion shaping results in over two octaves of spectral broadening with coherent DW–SHG spectral overlap, supporting robust *f*–2*f* detection. Beyond spectral expansion, the SCG process was shown to be highly tunable with respect to pump energy and input polarization, with a clear DW onset threshold around 102 pJ and a strong preference for TE-like polarization due to enhanced mode confinement and nonlinear overlap. By tracking the spectral co-evolution of DW and SHG across power levels, we directly verified the onset of spectral mixing essential for $f_{CEO}$ generation. A carrier-envelope offset beat note with 34 dB signal-to-noise ratio was achieved without optical amplification, and its strength was controllably tuned. To the best of our knowledge, this is the first time of amplification free $f_{CEO}$ detection with SNR~34 dB at 100 kHz RBW. These results position parabolic tapering in TFLN as a powerful and scalable approach to dispersion-managed nonlinear photonics, enabling compact, low-power, and fully integrated f–2f comb stabilization. This work lays the groundwork for next-generation portable optical clocks, chip-scale frequency synthesizers, and quantum photonic systems, and offers a broadly applicable method for nonlinear control in hybrid integrated photonics where spatial dispersion engineering is key.

**Funding.** National Key Research and Development Program of China (2024YFB2807400, 2024YFB2808600); National Natural Science Foundation of China (62305214, 62405035); Sichuan Science and Technology Program (No. 2023NSFSC0453); Fundamental Research Funds for the Central Universities (2024IAIS-QN018, 2024CDJYXTD-004); New Chongqing YC Project (CSTB2024YCJH-KYXM0105).

**Acknowledgment.** The facilities used for device fabrication were supported by the ShanghaiTech Material Device Lab (SMDL).

**Disclosures.** The authors declare no conflicts of interest.

**Data availability.** The data that support the findings of this study are available from the corresponding author upon reasonable request.